\documentclass[12pt,preprint]{aastex}
\shortauthors{Hsia et al.} 
\shorttitle{Study of Planetary Nebula NGC 6644}

\received{}
\usepackage{graphicx}
\graphicspath{{converted_graphics/}}
\begin{document}

\title{An Optical-Infrared Study of the Young Multipolar Planetary Nebula NGC 6644}

\author{Chih Hao Hsia, Sun Kwok, Yong Zhang}
\affil{Department of Physics, University of Hong Kong, Pokfulam Road, Hong Kong, China}
\email{xiazh@hku.hk; sunkwok@hku.hk; zhangy96@hku.hk}

\author{Nico Koning}
\affil{Department of Physics \& Astronomy, University of Calgary, Calgary, Alberta, Canada}
\email{nkoning@iras.ucalgary.ca}

\and

\author{Kevin Volk}
\affil{Space Telescope Science Institute, Baltimore, MD 21218}
\email{volk@stsci.edu}

\begin{abstract}

High-resolution HST imaging of the compact planetary nebula NGC 6644 has revealed two pairs of bipolar lobes and a central ring lying close to the plane of the sky.  From mid-infrared imaging obtained with the Gemini Telescope, we have found a dust torus which is oriented nearly perpendicular to one pair of the lobes.  We suggest that NGC 6644 is a multipolar nebula and have constructed a 3-D model which allows the visualization of the object from different lines of sight.  These results suggest that NGC 6644 may have similar intrinsic structures as other multipolar nebulae and the phenomenon of multipolar nebulosity may be more common than previously believed.

\end{abstract}

\keywords{ ISM: general --- planetary nebulae: individual 
(NGC 6644) --- stars: AGB}

\section{Introduction}

Although planetary nebulae (PNs) are usually assumed to have a simple spherical geometry, the actual morphologies of PNs have diverse forms.  When deeper, narrow-band observations are taken, PNs often reveal outer structures such as lobes and haloes.  While it has been recognized for some time that there are many examples of planetary nebulae with prominent bipolar structures (e.g., NGC 2346, NGC 6302), an increasing number of nebulae with multiple lobe structures have been found (e.g., NGC 2440, NGC 6072).  The origin of such structures and the physical mechanisms leading to their formation are yet to be understood.

Since PNs are made up of ionized, molecular, and dust components, a comprehensive understanding of the dynamical processes that create the observed morphological structures can only be obtained with multiwavelength observations.  With modern 8-m class telescopes equipped with high-resolution mid-infrared imaging capabilities, the distribution of the dust component can be compared to the distribution of the ionized gas component with comparable resolution.  In particular, the infrared observations will allow us to search for the existence of a high-density torus, and determine what role the torus play in the collimation of observed optical outflows.

The PN NGC 6644 (PNG 008.3-07.3, Hen 2-408, VV 188, IRAS 18295-2510) was discovered by \citet{hubble21}.
Based on early ground-based imaging, this compact nebula has been  classified as an elliptical PN by \citet{stan93}. It has a large heliocentric radial velocity of 205 km s$^{-1}$, 
and is suggested as a PN in the Galactic Bulge \citep{bea99}. However, \citet{gorny04} argued that this object is not a Galactic Bulge PN based on their spectroscopic study. 
Although NGC 6644 has a relatively high surface brightness, it has received 
very little attention in the literature. There are only three entries in ADS on the object after 
a search of keywords in the abstract, and the papers are mostly related to abundance studies of the nebula. 
\citet{aller88} proposed that the object was made from a much less metal-rich mixture than the Sun.

In this paper, we present results of our morphological study for NGC 6644 based on 
optical and infrared images taken with {\it HST} and {\it Gemini Telescope}, respectively. These images 
reveal for the first time unique morphologies and properties of the system. The observations and data reductions are described in $\S$ 2. In $\S$ 3, we present the results of 
imaging and spectroscopy in the visible and mid-infrared for this nebula. A discussion of the results, and 
a comparison of the observations with a three-dimensional model are presented in $\S$ 4. Finally, a 
conclusion is given in $\S$ 5.

\section{Observations and Data Reduction}

\subsection{HST Narrow-band Imaging}

Our {\it HST} data on NGC 6644 were retrieved from the Space Telescope Science Archive.
The images of NGC 6644 were obtained under program 8345
(PI: R. Sahai) using the Wide Field Planetary Camera 2 (WFPC 2) on {\it HST}. The object was observed with the  Planetary Camera (PC) on February 26, 2000, which provides
a 36.$\arcsec$8 $\times$ 36.$\arcsec$8 field of view (FOV) at a spatial resolution of 0.045$\arcsec$ pixel$^{-1}$.  The actual observations were made with different exposure times 
(from 20 s to 400 s) to allow for the imaging of both the bright central region and the faint outer parts. 
The data were processed through the {\it HST} pipeline calibration.  Standard bias subtraction and flat-field correction were performed. Data were taken in two-step dithered positions to enhance spatial sampling and cosmic rays removal by using the task {\bf crrej} in the STSDAS package of IRAF. The processed   F656N (H$\alpha$) observation ($\lambda_{p}$ = 6564 \AA, $\triangle\lambda$ = 22 \AA) with a total exposure time of 1120 s is shown in Fig.~\ref{hst}. 

\subsection{Gemini T-ReCS observations} \label{trecs}

The imaging observations were made with the Thermal-Region Camera Spectrograph (T-ReCS) on the {\it Gemini South 
Telescope} under program GS-2007A-DD-9 on 2007 October 10. The field of view of the detector is 28.$\arcsec$8 $\times$ 21.$\arcsec$6 with a pixel scale of 0.$\arcsec$09. The background emission from the sky and telescope were removed with chopping and nodding during the observations. NGC 6644 was imaged with three 
narrow/medium band filters: [S~{\sc iv}] ($\lambda_{c}$ = 10.52 $\micron$, $\triangle$ $\lambda$ = 0.17 $\micron$), PAH-2 ($\lambda_{c}$ = 11.30 $\micron$, $\triangle$ $\lambda$ = 0.61 $\micron$), Si-5 ($\lambda_{c}$ = 11.66 $\micron$, $\triangle$ $\lambda$ = 1.13 $\micron$), and the broad $Q_a$ band filter 
($\lambda_{c}$ = 18.30 $\micron$, $\triangle$ $\lambda$ = 1.38 $\micron$). The total on-source exposure times for the [S~{\sc iv}], PAH-2, Si-5, and $Q_a$ filters are 455.1, 461.5, 238.9, and 238.9 s, respectively. The measurements were made under good (20 percentile) sky conditions, and the mid-IR images are diffraction limited.
The measured FWHM values for the standard stars are 0.375, 0.382, 0.427, and 0.552 \arcsec for the [S~{\sc iv}], PAH-2, Si-5, and $Q_a$ filters respectively.
Photometric standard stars were observed for the purpose of flux calibration using the spectrophotometric fluxes for the standard stars from \citet{cohen99}.  The journal of observations is summarized in Table 1.

\subsection{Spitzer IRS Spectrum}
\label{sec_irs}

The mid-infrared spectra of NGC 6644 were obtained by the IRS \citep{houck04} through the observation
program 3633 (PI: M. Bobrowsky) with Astronomical Observation Request (AOR) key of 11334400. The measurements
were observed using the Short-Low (SL) module (5.2 $\micron$ - 14.5 $\micron$), Short High (SH) module (9.9 $\micron$ - 19.6 $\micron$), and the Long High (LH) module (18.7 $\micron$ - 37.2 $\micron$) with spectral dispersions of $R\sim$ 600. 
The diaphragm sizes are 3$\farcs$6 $\times$ 57$\arcsec$, 
4$\farcs$7 $\times$ 11$\farcs$3, and 11$\farcs$1 $\times$ 22$\farcs$3  in SL, SH and LH modules respectively. 
The total integration time of IRS observation is 347 s.  The IRS measurement of NGC 6644 was not positioned at the central star but on the outer lobe. An overlay of the IRS apertures on the optical image of the nebula is
shown in Fig.~\ref{aperture}.

Data were reduced starting with basic calibrated data (BCD) from the Spitzer Science Center's
pipeline version s18.7 and were run through the {\bf IRSCLEAN} program to remove rogue pixels.
Next the {\bf SMART} analysis package \citep{higdon04} was used to extract the spectra.
A final spectrum was produced  using the combined IRS observations to improve the signal-to-noise
ratio (S/N).  
Since the IRS spectrum in the short wavelength range is taken from the outside of the nebula, some scaling is needed for the shorter wavelength observations.  In Fig.~\ref{volk}, we scale the IRS SH and SL observations by factors of 1.29 and 1.625 respectively and are able to obtain a smooth spectrum.  When we compare these adjusted flux levels to the photometric measurements obtained from the 4 Gemini T-ReCS filters, we find good agreement (Fig.~\ref{volk}).  Further confidence of this adjustment is found by the agreement between the spectra and the IRAS 12 and 25 $\mu$m fluxes.  
Also plotted in Fig.~\ref{volk} are the 4 Gemini filter transmission functions.  We can see that the [S~{\sc iv}] and PAH-2 filters have most of their fluxes contributed by the [S~{\sc iv}] line and the 11.3 $\mu$m aromatic infrared band (AIB) respectively.  The Si-5 filter mainly takes in fluxes from the 12 $\mu$m plateau emission feature and the $Q_a$ filter measures the general dust continuum.  Although the $Q_a$ filter covers the [S~{\sc iii}] line, we estimate the line contribution to the filter flux to be $\sim$8\% and should not affect the observed morphology of the dust continuum.

\section{Results}

\subsection{Lobes and Halo}\label{lobes}

\begin{figure}
\epsscale{1.1}
\plotone{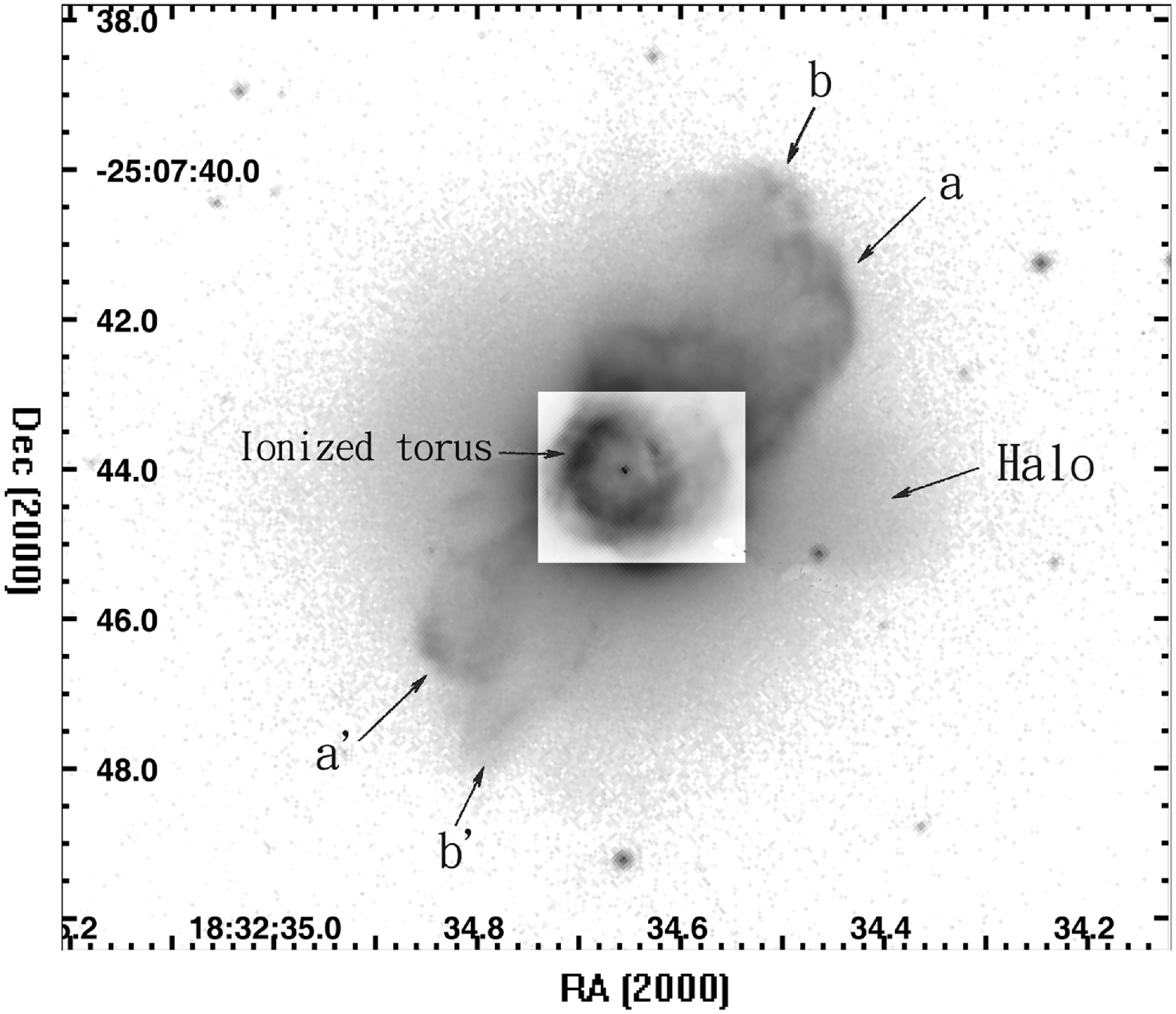}
\caption{{\it HST} H$\alpha$ image of NGC 6644 showing various morphological structures.  
The inner (central) and outer regions are displayed at different levels to better show the structures.
The two bipolar lobes are marked as $a-a^\prime$ and $b-b^\prime$ and the central ring structure is marked as ``ionized torus''.  The nebulosity associated with the ring (marked as ``c'') could be the projection of another pair of lobes aligned almost along the line of sight.   An extensive halo (marked) can also be seen. \label{hst}}
\end{figure}

\begin{figure}
\plotone{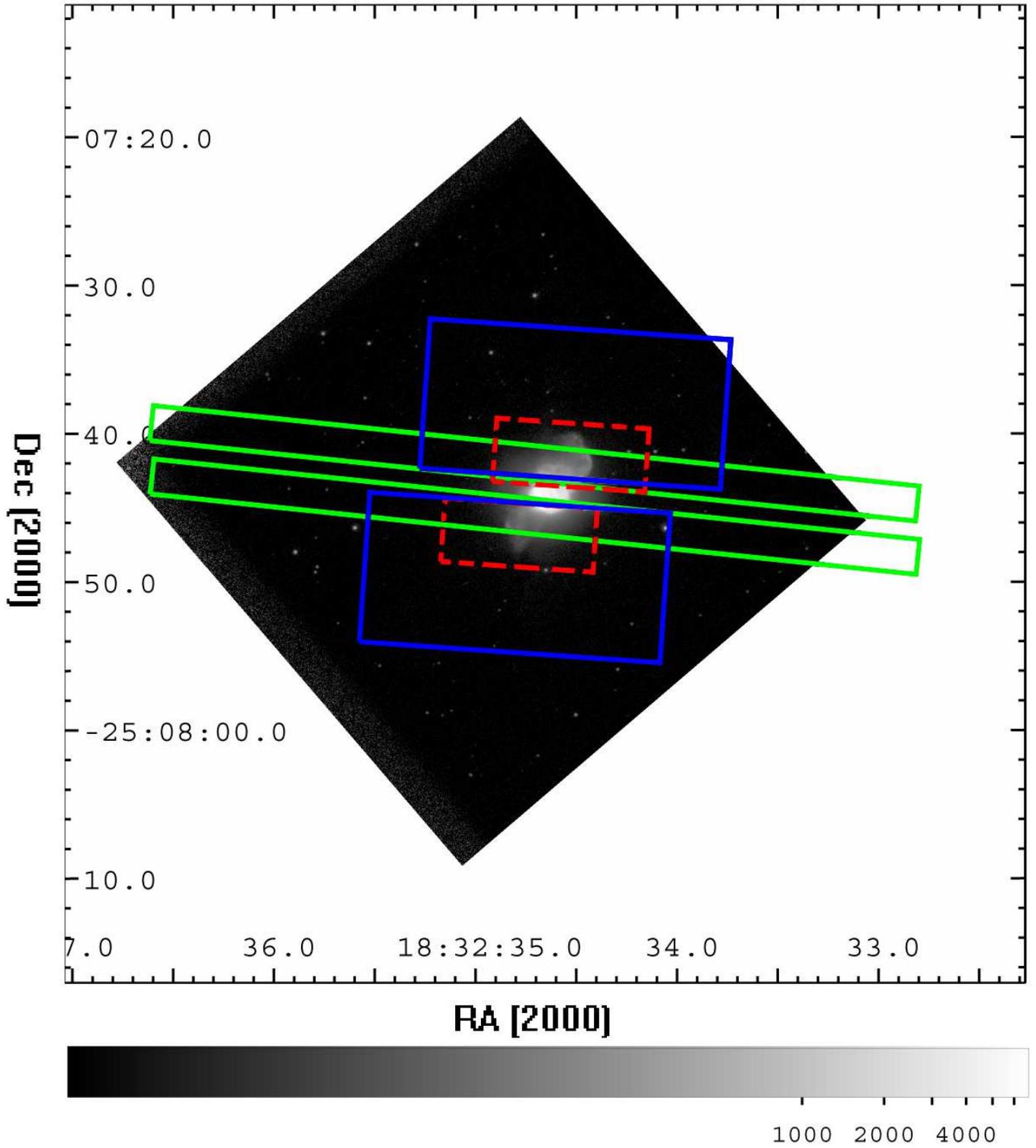}
\caption{Apertures of the {\it Spitzer} IRS observations overlaid on the {\it HST} image of NGC 6644.  The red, blue, and green boxes show the SH (10-20 $\mu$m), LH (19-37 $\mu$m) and SL (5.2-14.5 $\mu$m) observations respectively.}
\label{aperture}
\end{figure}

\begin{figure}
\plotone{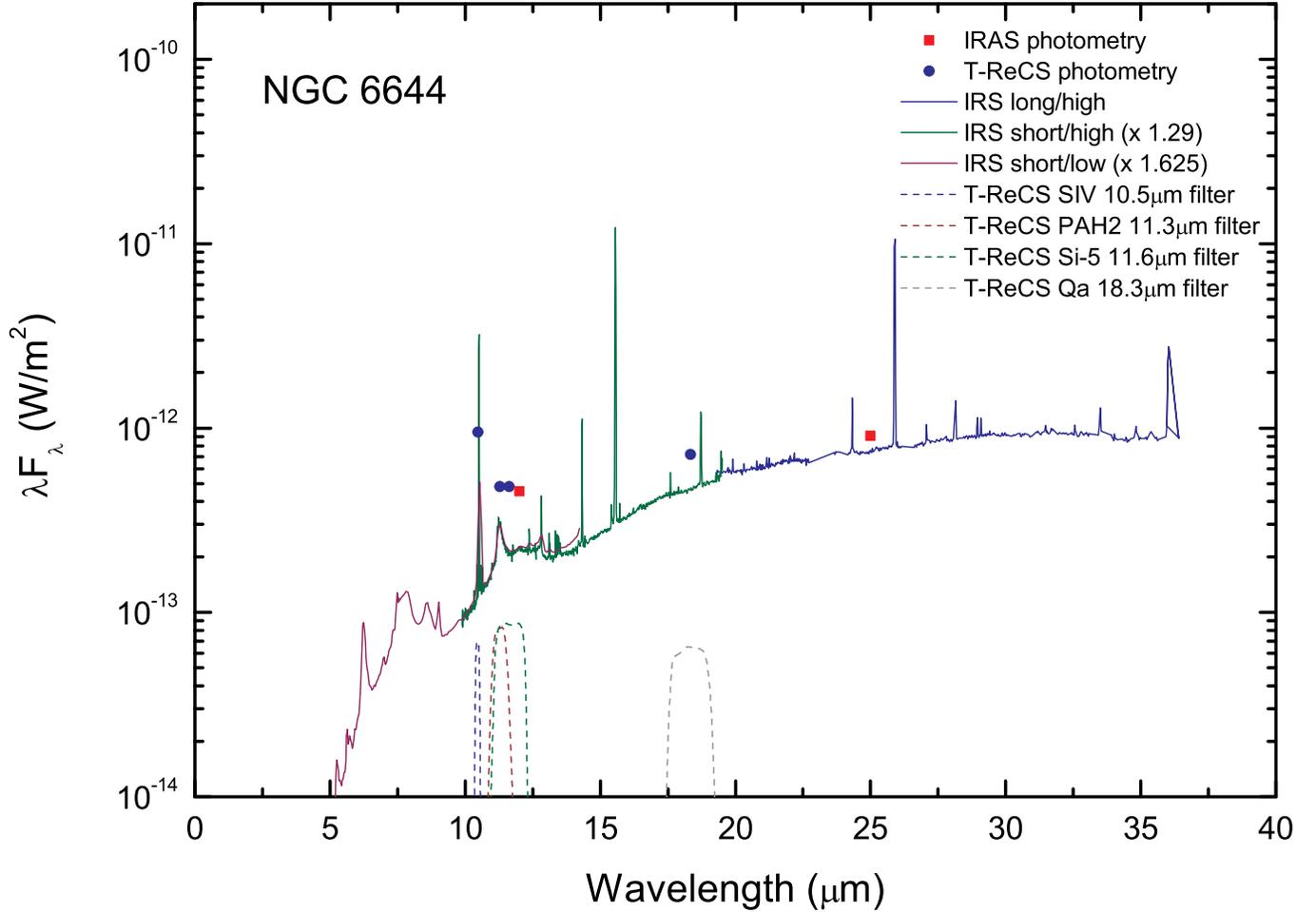}
\caption{Comparison between the Spitzer IRS spectra and the Gemini T-ReCS photometry.  The observations of the IRS SH and SL have been adjusted upward by factors of 1.29 and 1.625 respectively (see text).  The T-ReCS and IRAS photometric measurements are shown as circles and squares respectively. The Gemini T-ReCS filter transmission functions are plotted as dotted lines.}
\label{volk}
\end{figure}

\begin{figure}
\plotone{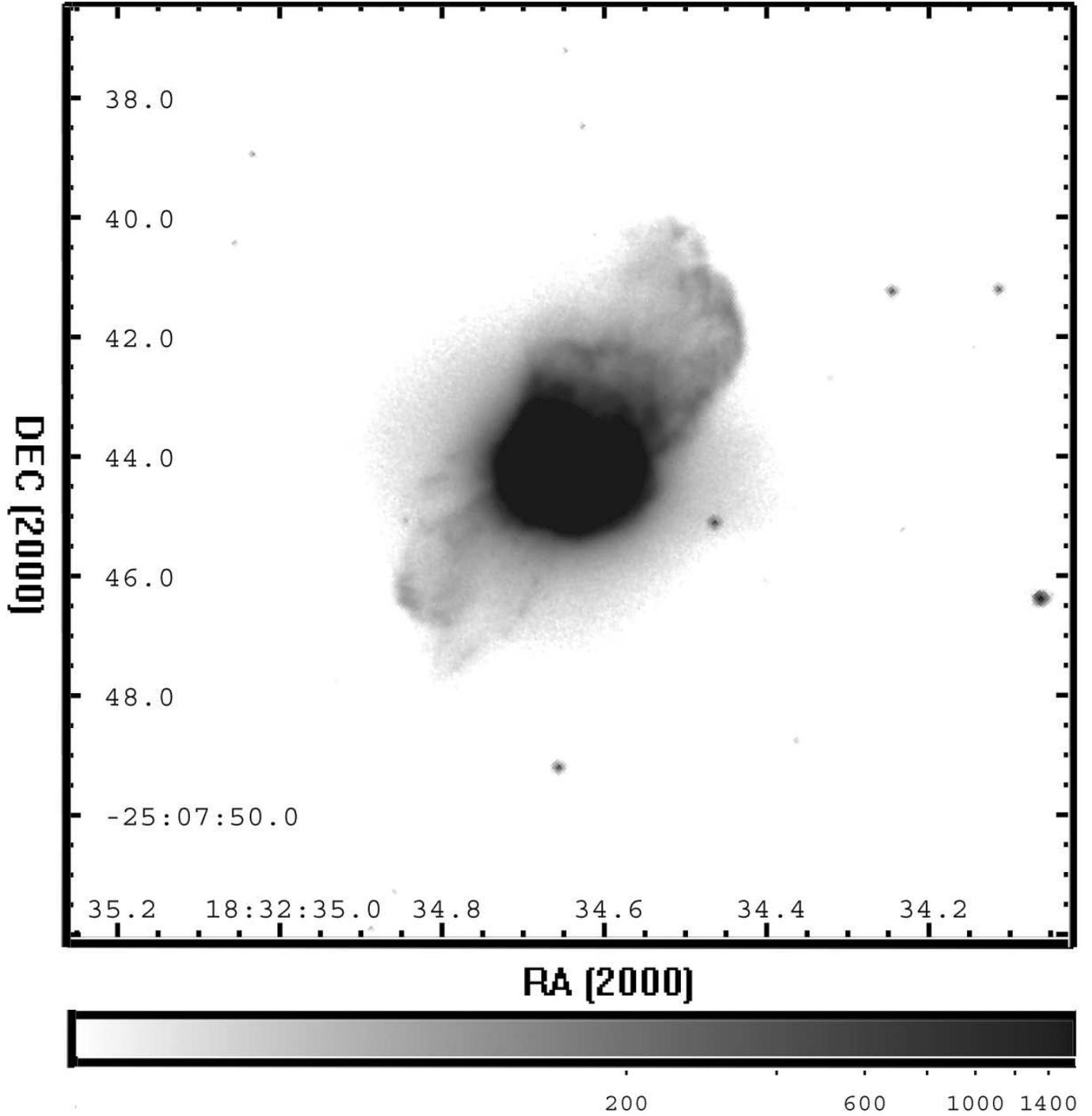}
\caption{The {\it HST} H$\alpha$ image of NGC 6644 with an intensity display setting to show the outer structures of the nebula.  The intensity display is on a logarithmic scale and the grey scale bar is given at the bottom in units of counts per pixel.}
\label{over}
\end{figure}

\begin{figure}
\plotone{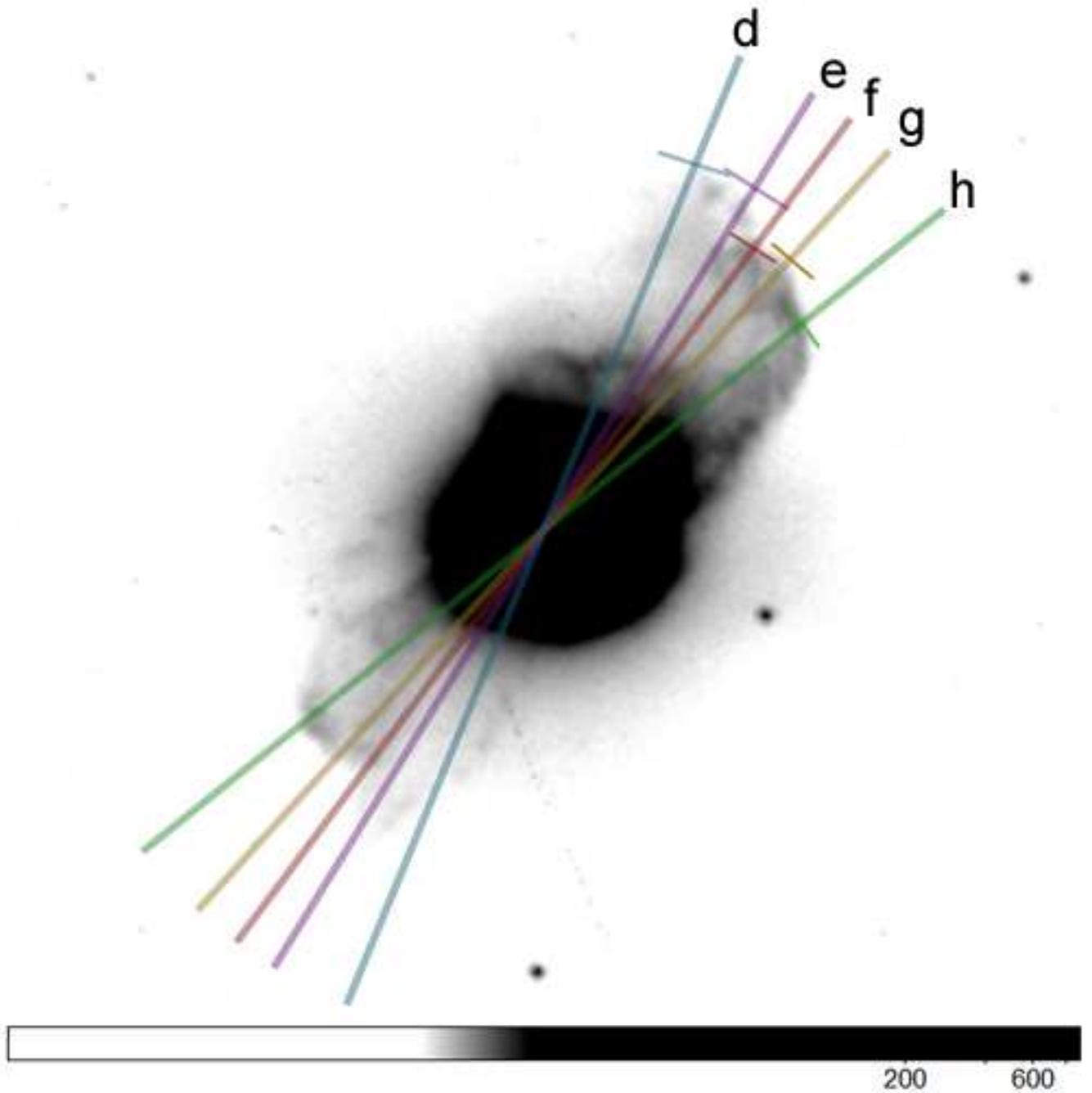}
\caption{A number of faint sub-lobes can be seen in the {\it HST} H$\alpha$ image of NGC 6644.  The caps of the lobes have been highlighted by bars for easy identification.  The lines perpendicular to the bars labeled as $d, e, f, g, h$ are inferred axes of the lobes.}
\label{nico2}
\end{figure}

\begin{figure}
\plotone{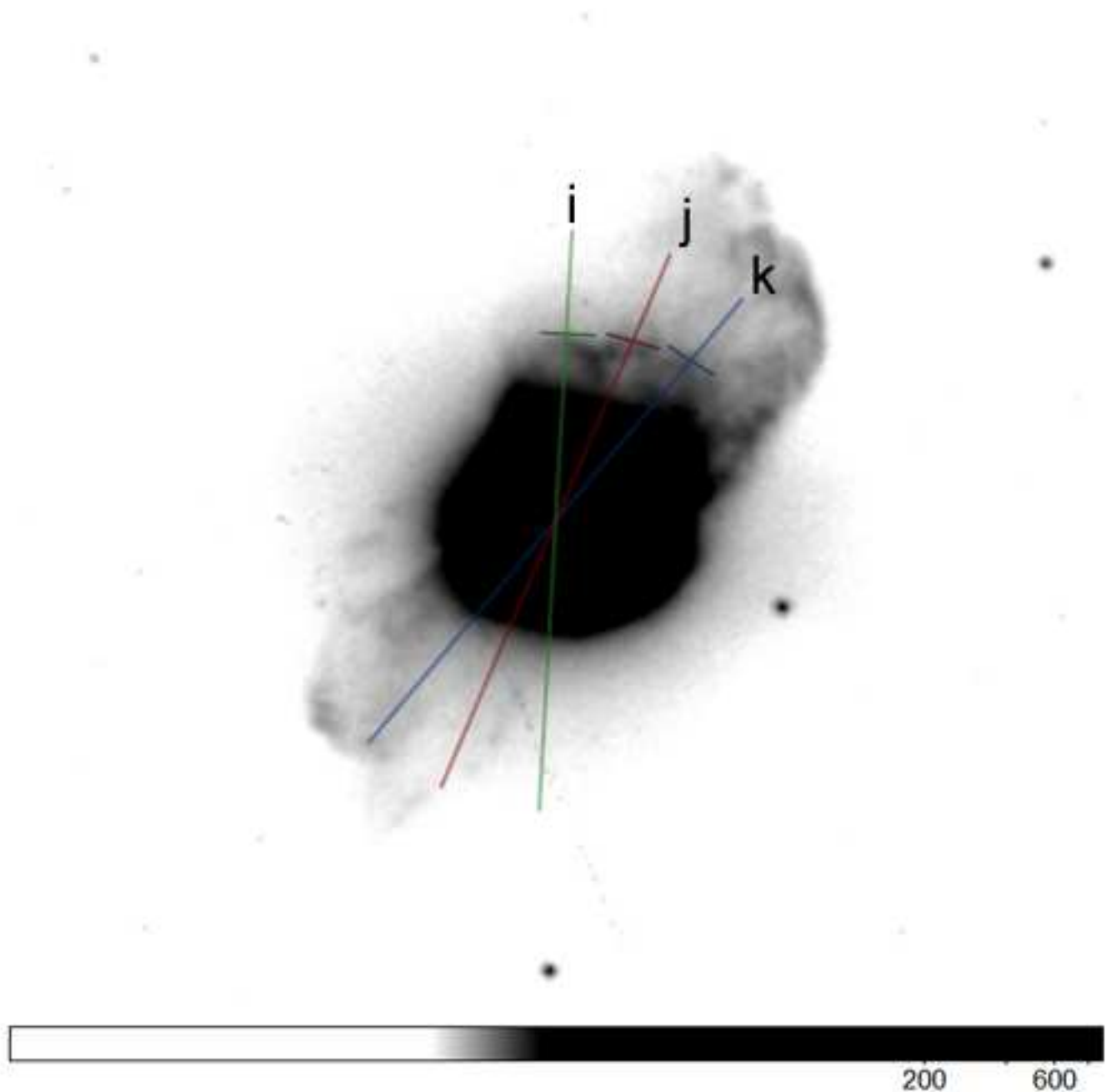}
\caption{Three inner lobes are marked on the {\it HST} H$\alpha$ image of NGC 6644.  The labels are as in Fig.~\ref{nico2}.}\label{nico3}
\end{figure}

The {\it HST} H$\alpha$ image of NGC 6644 reveals several morphological structures in the nebula (Fig.~\ref{hst}).  Easily discernible are two bipolar lobes (labeled as $a-a^\prime$ and $b-b^\prime$) 
along the approximate SE-NW direction.  A ring structure (labeled as ``ionized torus'') near the center can also be seen.    Beyond the bright ring, there are associated nebulosities (marked as ``$c$''), in particular in the western direction.  It is possible that these nebulosities represent the projection of a third pair of lobes which are  aligned nearly along the line of sight.  An extended diffuse structure (marked as ``halo'') can also be seen.
In Fig.~\ref{over}, we display the image with the central part of the nebula saturated, allowing the fainter outer structures to be seen.  There are apparently sub-structures in the nebula. In 
Fig.~\ref{nico2}, we mark a few possible sub-structures (which appear predominantly in the north) as $d, e, f, g, h$.
Closer to the core of NGC 6644 appears several more sub-substructures, again most obvious in the northern part of the image. These are labeled $i, j, k$ in Fig.~\ref{nico3}. 
The axes of these structures are defined by the caps of the lobes, location are marked by horizontal bars in Figs. ~\ref{nico2} and \ref{nico3}.  The orientation of the axis of $d$ is less well defined as the cap is not complete, but the others are reasonably well defined.  
What is interesting is that the axis $j$ seems to lie along the same orientation as $d$ , as well as the axes of $k$ and $g$.  
The fact that these sub-structures appear stronger and more evident in the northern half of the image may provide further evidence that the north is tilted towards us and therefore experiences less extinction than the south.

The symmetry axes of lobes $a$ and $b$ intersect approximately at the position of the central star. 
The two pairs of butterfly-shaped, closed-end bipolar lobes ($a-a^\prime$ and $b-b^\prime$) have symmetry axes that lie  at the position angles (PAs) of -43 $\pm$2$^\circ$ and -27$\pm$2$^\circ$, respectively. 
The lobes $a-a^\prime$ and $b-b^\prime$ have similar projected sizes on the sky.  Lobe $a-a^\prime$ has a projected length of $\sim$ 7.76$\arcsec$ and $b-b^\prime$ has a projected size of 
$\sim$ 8.81$\arcsec$. 
If we assume a distance of 2.5 kpc \citep{cahn92} 
the physical size of the total extent of the $b-b^\prime$ lobe is $\sim$ 0.35 $\sec \theta$ light yr,  where $\theta$ is the inclination angle. For an expansion velocity of 100 km s$^{-1}$, the kinematic age of this lobe is $\sim$ 521 yrs if the lobes lie close to the plane of the sky. From this kinematic age,
NGC 6644 can be considered as a young PN.

For the central ring (``ionized torus''), its major axis is found to have an angular orientation of  PA $\sim$ -55$\pm$3$^\circ$. Similar structures have been found in other multipolar PNs such as the two bright partial rings seen in the waist of He 2-47 \citep[Fig.~1b]{sahai00b}.  
The eastern segment of the ring in NGC 6644 is brighter than the western side, as evident in both the H$\alpha$ image (Fig.~\ref{hst}) and in the [S~{\sc iv}] image in Fig.~\ref{gemini}.
This suggests that the eastern segment is closer to us, whereas the western side suffers from extinction from intervening dust in the dust torus.
 
\begin{figure}
\plotone{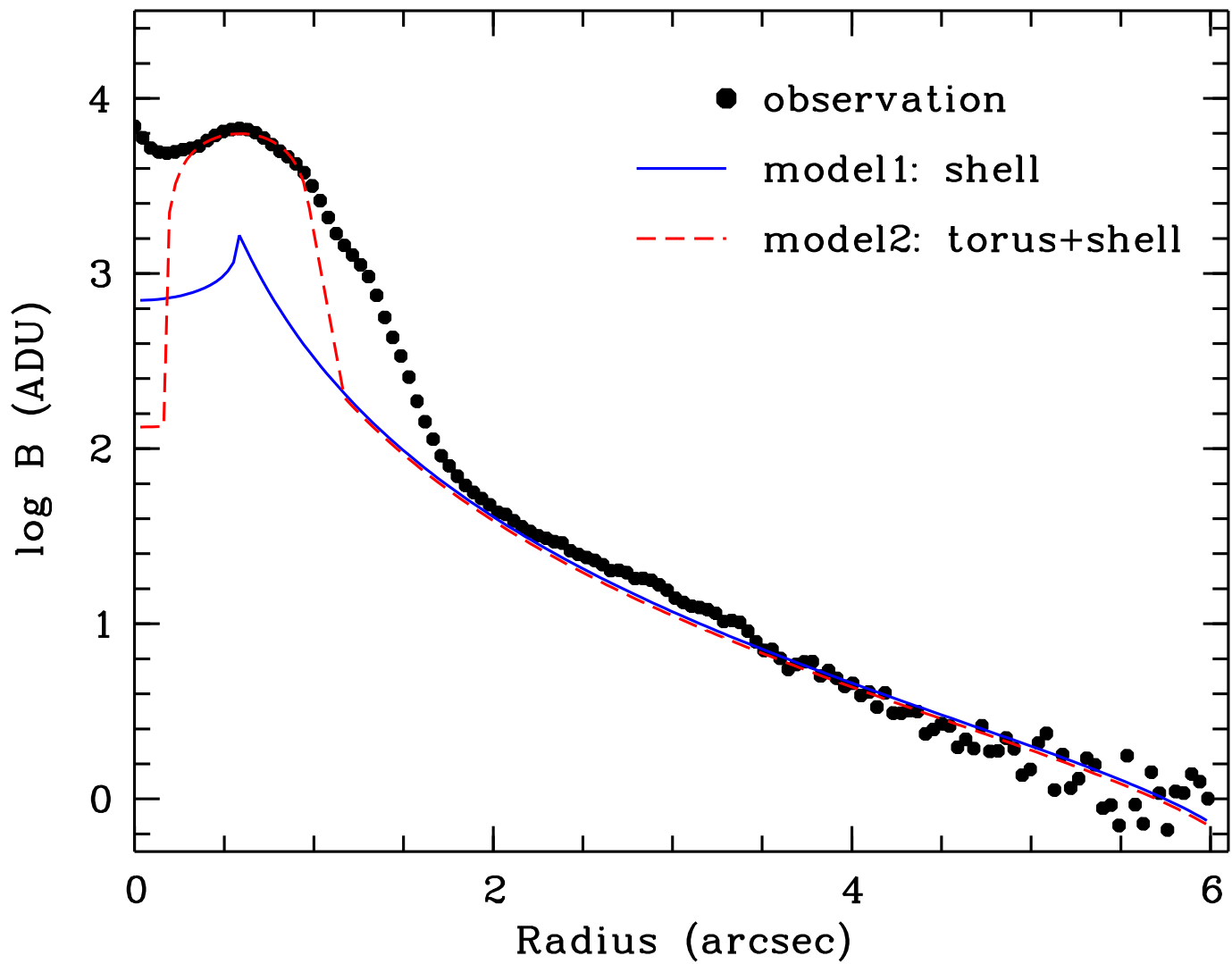}
\caption{Profile of surface brightness distribution of  H$\alpha$ emission in NGC 6644 averaged over all angles except the regions occupied by the lobes $a-a^\prime$ and $b-b^\prime$.  The halo can be fitted by a density law of  r$^{-2}$ (model in solid line) but the intensities in the inner regions require contribution from a separate component (dashed line). See text for details.  \label{profile}}
\end{figure}

This bright multipolar nebula is surrounded by a faint, extensive halo with a well-defined spherical shape. 
The extended halo emission at H$\alpha$ can be seen in an averaged radial surface brightness profile 
(with the two bipolar lobes excluded) shown in Fig.~\ref{profile}. This profile is constructed from measurements of 5\degr\ intervals between PA=25\degr-85\degr and PA=205\degr-265\degr, after the removal of all field stars.
The intensity of the nebula within 1.$\arcsec$4 is dominated by the ionized torus and the angular radius of the faint halo is estimated to be larger than 4\arcsec.  The surface brightness of the extended part at radius 4\arcsec\  relative to that in the peak is $\sim$ 10$^{-3}$.
At a distance of 2.5 kpc, the physical size of the halo is 0.15 light-yr, with a kinematic age of  
$\sim$ 3$\times$ 10$^{3}$ yr, if the expansion velocity of the halo is 15 km s$^{-1}$.

An intensity profile of the nebula (minus the lobes) averaged over all angles is shown in Fig.~\ref{profile}.  Also shown is a fit of the profile by a halo density distribution of $r^{-\alpha}$, where $r$ is the
radial distance from the central star, with an inner and outer radii of $R_{\rm in,out}=0.585$ and 6.5 arc sec respectively. For the modeling, we have assumed that H$\alpha$ is optically thin and its intensity integrated along each line of sight is proportional to the emission measure,  $EM(p)=\int_p n^2(r) d\ell$, where $n(r)$ is the density of hydrogen ion,  $p$ is the angular distance from the central star, and $d\ell$ is the path length element along the line of sight. Therefore, under the assumption that the halo is a spherical shell, we have

\begin{equation}
EM(p)=2{n^2(R_{\rm in})}{R_{\rm in}^{2\alpha}}p^{1-2\alpha}\int^{\theta_2}_{\theta_1}
( \cos\theta )^{2\alpha-2} d\theta,
\end{equation}
where $\theta_1=\cos^{-1}(p/R_{\rm in})$ and 0 for
$R_{\rm in}\le p \le R_{\rm out}$ and $p<R_{\rm in}$ respectively, and $\theta_2=\cos^{-1}(p/R_{\rm out})$.
For a stellar wind with a constant mass-loss rate and expansion velocity, $\alpha$ has a value of 2 and the density distribution varies as  $n(r)=n(R_{\rm in})(r/R_{\rm in})^{-2}$.  While the model provides a reasonably good fit to the halo, it fails in the inner region.

In a second model, we introduce a face-on torus with a cross-section radius of $R_{\rm t}=0\farcs4$ and  a homogeneous density distribution of $n_{\rm t}$.
For the torus, we have
\begin{equation}
EM(p)=2n_{\rm t}^2\sqrt{R_{\rm t}^2-(p-R_{\rm c})^2},
\end{equation}
where $R_c=0\farcs585$ is the distance from the circle center of the torus to the central star.  The torus is embedded in a halo with a inner radius of 1 arc sec. This  addition makes an improved fit to the intensity profile in the central region. The remaining excess in emission between 1 and 2 arc sec (as seen in Fig.~\ref{profile}) could be due to
emission from the nearly pole-on lobes.

In the model, we do not consider the effect of extinction. Due to the presence of dust torus (see below), the central region may have a higher extinction than the halo. If this is the case, $EM(p)$ given in eq. (2) could be underestimated and the actual intensity of the inner region would be even higher.  This will further strengthen the need for a face-on torus.

\subsection{The dust torus}

\begin{figure}
\plotone{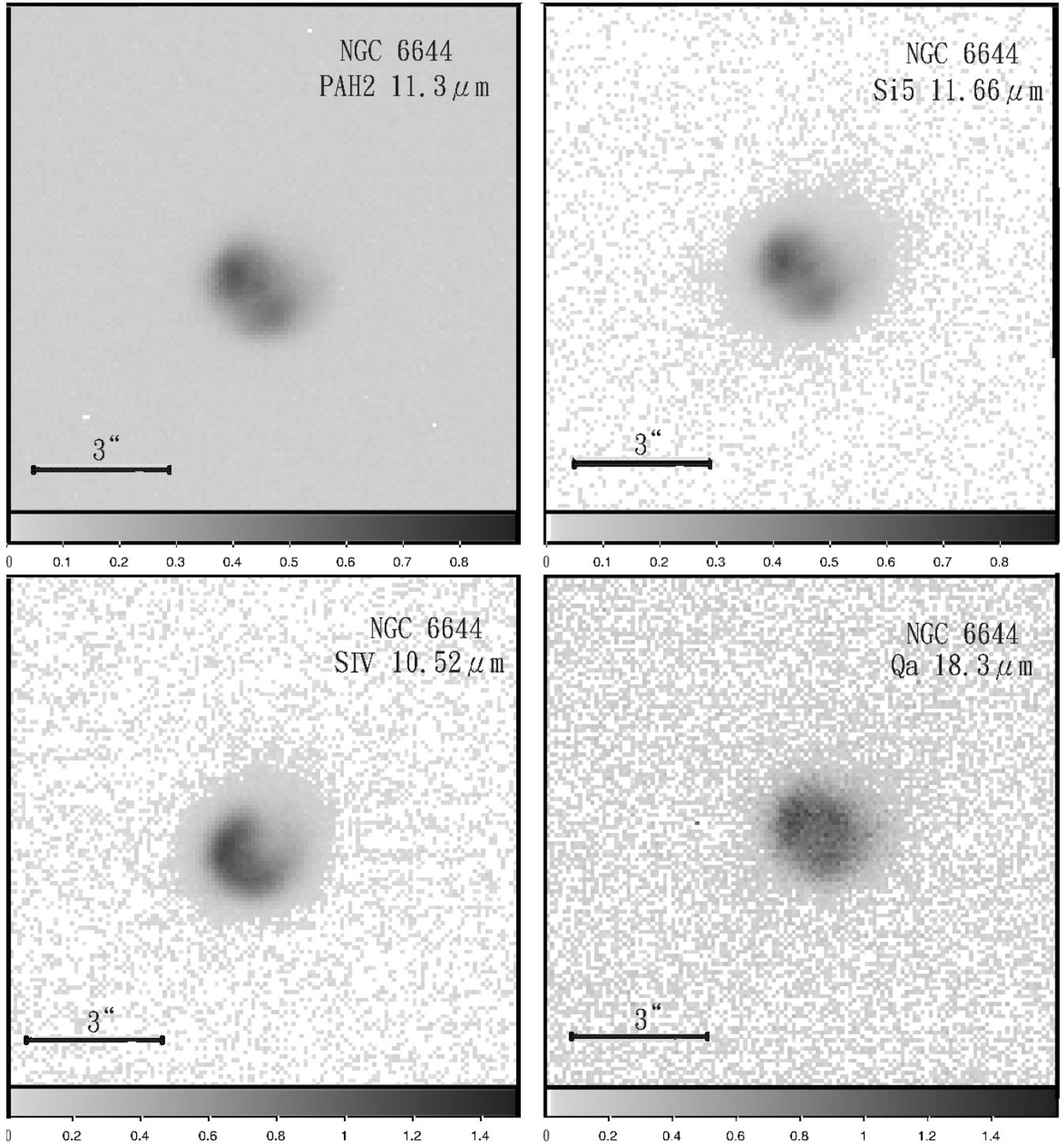}
\caption{Flux-calibrated {\it Gemini} T-ReCS images of NGC 6644. The four panels give respectively the 
PAH-2 11.3 $\micron$ image (upper left), the Si5 11.66 $\micron$ image (upper right), [\ion{S}{4}] 10.52 
$\micron$ narrow-band image (lower left), and $Q_a$ 18.3 $\micron$ medium-band image (lower right). North is up and East is to the left. The intensity display is on a linear scale and the grey scale bar is given at the bottom in units of counts per pixel.\label{gemini}}
\end{figure}

\begin{figure}
\plotone{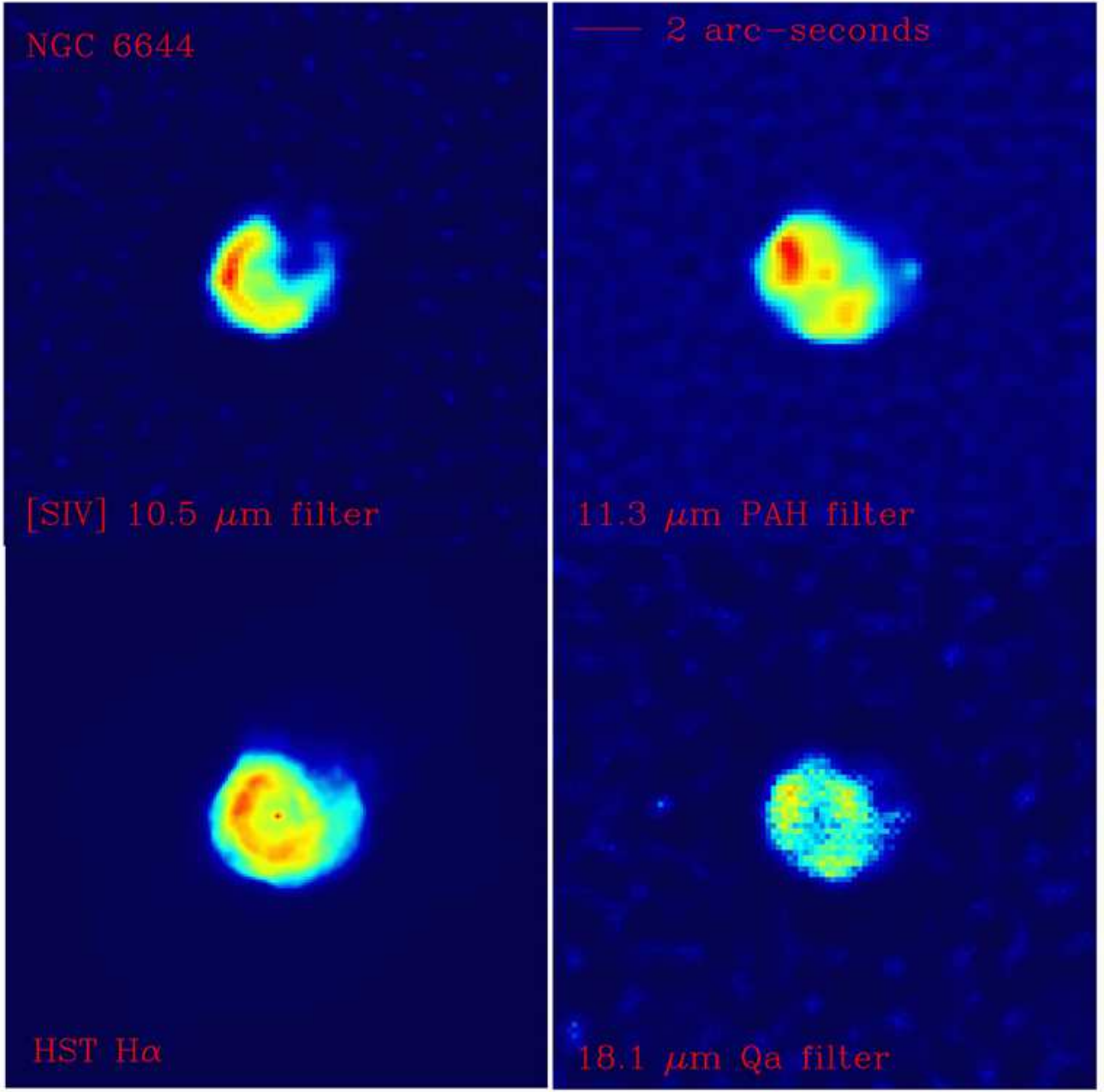}
\caption{Comparison between the  {\it HST} H $\alpha$ image (lower left) and the flux-calibrated, deconvolved, false-color {\it Gemini} T-ReCS images displayed on the same pixel scale.  The Gemini [\ion{S}{4}] 10.52 $\micron$ narrow-band, PAH-2 11.3 $\micron$ narrow-band, and the  $Q_a$ 18.3 $\micron$ medium-band images are given in the upper left, upper right, and lower right panels respectively.}
\label{decon}
\end{figure}

In Fig.~\ref{gemini} we show the 4-band T-ReCS (PAH-2, Si-5, [S~{\sc iv}], and $Q_a$) flux-calibrated images in units of Jy per square arcsecond. 
The T-ReCS images of NGC 6644 were each deconvolved using the observations of the standard star HD 175775 taken immediately after the science observations.  The ``lucy'' task in the ``analysis'' group of the ``stsci'' external IRAF package was used for the deconvolutions.  For each filter a background-subtracted, normalized PSF image of size 51 by 51 pixels was created from the standard star observation, and this was used as the kernel for the deconvolution. The deconvolution was iterated 25 times to produce the final deconvolved images.  It was found that most of the improvement in the images took place within the first 5 to 10 iterations.  Under the seeing conditions reported in section \ref{trecs}, the Lucy deconvolution is robust to small changes in the FWHM of the PSF.   
The deconvolved images are shown in Fig.~\ref{decon} together with the H$\alpha$ image on the same scale.

From Figs.~\ref{gemini} and \ref{decon}, we can clearly see that there is a spatially resolved  structure in the central region of NGC 6644.  The [S~{\sc iv}] image shows a partial ring structure similar to that seen in the central region in the H$\alpha$ image, suggesting that they both trace the ionized gas distribution.  The Si5 and PAH2 images are different from the [S~{\sc iv}] image, showing a double-peaked structure.   The $Q_a$ image shows a more diffuse structure than the other 3 images.  The overall maximum angular extent of the emission regions in the four images are similar.  A more detailed comparison between the {\it HST} H$\alpha$ image with the {\it Gemini} [S~{\sc iv}] image shows that the ring in the [S~{\sc iv}] image is less complete than its counterpart in the H$\alpha$ image (Fig.~\ref{decon}).  There are two factors that can cause a difference between the two images.  The [S~{\sc iv}] image includes contributions from both the [S~{\sc iv}] line and dust continuum, and therefore should reflect the morphology of the ionized torus as well as the oblique dust torus.  Since the dust torus (see Fig.~\ref{gemini}) is brighter in the eastern side, the [S~{\sc iv}] image should therefore appear as a less complete ring than in the H$\alpha$ image.  If there is also effect of local extinction that causes the eastern side to be brighter (see section \ref{lobes}), then the [S~{\sc iv}] image should suffer from less extinction and therefore should show a more complete ring.  This effect is apparently not as important.

If the double-peaked structure in the PAH2 and Si5 filter images is interpreted as an oblique torus, the size of the torus has dimensions of 28 $\times$ 15 pixels with the major axis lying along PA $\sim$ -44$\pm$3$^\circ$.  This translates to a major axis size of 2$\farcs$52 and a minor axis size of 1$\farcs$35. Assuming that this ellipse is a projection of a circle on the sky, an angle of tilt of the dust torus is approximately 32$^\circ$ (with 90$^\circ$ being pole on).  The fact that this torus is not obvious in the optical image is probably because the dust torus is optically thick to the UV photons and the volume of the torus is not ionized.
  
In Table 3, we compare the fluxes observed in the four band T-ReCS observations and fluxes from the IRS observations.  Columns 1 and 2 list the observed filter names and central wavelengths of these filters. The corresponding band width for each filter is given in column 3. The measured integrated fluxes from the T-ReCS observation are given in column 4. 
The simulated T-ReCS in-band fluxes derived from the IRS spectrum and the filter profiles of the 4 bands are given in column 5.  There is general consistency between the Gemini and IRS measurements with the discrepancies being due to varying sky conditions and uncertainty of photometric flux calibrations.  These uncertainties are higher in the $Q_a$ band.

\subsection{Infrared features}

\begin{figure}
\epsscale{0.8}
\plotone{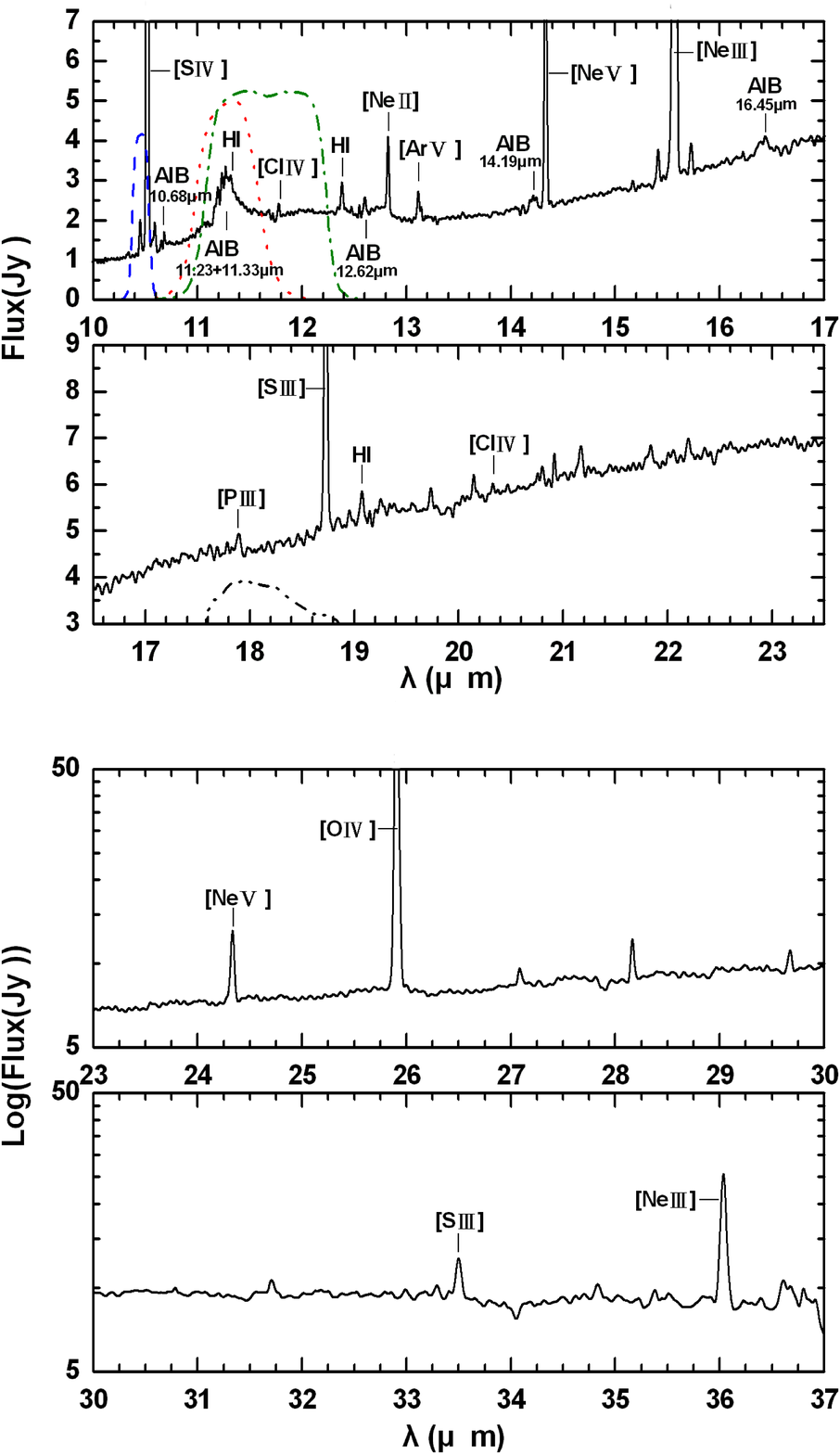}
\caption{{\it Spitzer} IRS spectrum of NGC 6644 in wavelength range from 10 $\micron$ to 37 
$\micron$. The emission lines and AIB features are marked.  The Gemini T-ReCS filter transmission functions are plotted as dotted lines.\label{irs_spectrum}}
\end{figure}

The combined IRS spectrum of NGC 6644  shown in  (Fig.~\ref{volk}) shows a strong infrared continuum due to dust emission peaking at $\sim$ 30 $\micron$.  On top of this continuum are a number of emission lines typical of the spectrum of PNs.
Among the strongest emission lines detected are the  fine structure lines of [\ion{S}{4}]  at 10.51 $\micron$, [\ion{O}{4}] at 25.88 $\micron$, [\ion{Ne}{3}] at 15.55 and 36.01 $\micron$, and weaker H recombination lines \ion{H}{1} (7-6) at 12.37 $\mu$m and \ion{H}{1} (8-7) at 19.07 $\micron$ (Fig.~\ref{irs_spectrum}).
The measured emission line fluxes in the spectrum are given in Table 2. The first two columns in Table 2 contain the central wavelengths of the emissions and line identifications, respectively. Column 3 gives the observed fluxes measured using the Gaussian fitting routine. 
Using the fine-structure lines [S~{\sc iii}] and [Ne~{\sc iii}], we can derive the electron density.  From the [S~{\sc iii}]  18.7 $\mu$m to 33.4 $\mu$m and [Ne~{\sc iii}] 15.6 $\mu$m to 36 $\mu$m line ratios, we derive $\log n_e=3.8$ and 3.4 cm$^{-3}$ respectively.
The last column gives the total flux of the entire nebula by scaling the observed flux by a factor, derived from the average theoretical flux ratio between infrared \ion{H}{1} lines (\ion{H}{1} (7-6), \ion{H}{1} (8-7)) and H$\beta$ emission based on Case B 
recombination line theory.  Assuming an electron temperature of $T_{e}$= 12,000 K and an electron density 
of $n_{e}$ = 7,200 cm$^{-3}$ \citep{shaw89}, the average theoretical ratios are then combined to derive 
the scaling factor of 1.01 for the IRS spectrum.

Also present is the strong, broad AIB at 11.3 $\mu$m.  This feature is due to the C-H out-of-plane bending of aromatic compounds and is commonly seen in carbon-rich PNs and proto-PNs \citep{kwok99}.  Several other broad features at 12.6, 14.2, 16.5 $\mu$m are also likely to be AIB features.  A strong, broad 12 $\mu$m plateau emission feature, attributed to out-of-plane bending modes of a mixture of aliphatic groups attached to the aromatic rings \citep{kwok01}, can also be seen.
A recent study of AIB features in Galactic Bulge PNs using IRS data was presented by \citet{perea09}, in which they find AIB features commonly present in their sample.

\subsection{Spectral Energy Distribution}

Although PNs were first known for their optical emission-line characteristics, it was realized after the IRAS mission that a significant amount of the total energy output is emitted in the infrared region due to thermal emission by dust grains.  
A systematic investigation for spectral energy distributions (SEDs) of young PNs covering  the wavelength range  from ultraviolet (UV) to far-infrared was performed by  \citet{zh91}.
Depending on the stage of evolution, it is found that the photospheric, ionized gas, and dust components are the main contributors to the total observed flux. 
In constructing the SED of NGC 6644, we have made use of data in the astronomical data archives of both ground-based and space-based observations.

In the UV region, we have used observations from the {\it International Ultraviolet Explorer} (IUE) low
dispersion spectrograph. All {\it IUE} spectra of the object are produced at the IUE Data Analysis
Center (IUEDAC). Reduced data of this nebula are taken in co-added positions of the short wavelength 
prime (SWP) and long wavelength prime (LWP) spectra to enhance
the signal-to-noise ratio, respectively. The parts of this spectrum in SWP and LWP are then 
combined at 1975 \AA\ to produce a single, resultant spectrum.

A journal of the IUE data and {\it Spitzer} IRS AORkey of the nebula are given in Table 4.
In the infrared, the {\it Spitzer} IRS observations are used together with the photometric observations of {\it Infrared Astronomical Satellite} (IRAS).  The photometric measurements of NGC 6644 at the four wavelength bands of 12, 25, 60, and 100 $\micron$ are taken from \citet{tajitsu98}.

The B and V magnitudes of the central star of NGC 6644 are observed by \citet{shaw89}. Near-infrared magnitudes of I, J, H, and K/Ks observed by 
the {\it Deep Near-Infrared Southern Sky Survey} (DENIS) and the {\it Two Micron All
Sky Survey} (2MASS) are derived from DENIS database and \citet{ramos2005}.  
A summary of these archival data is given in Table 5.

After correcting the UV spectrum and optical measurements 
by an extinction value of 0.41 \citep{shaw89} (which accounts for both circumstellar and interstellar extinction), we fitted the emerging flux ($F_\lambda(total)$) by a three-component model including a hot central star, a warm gaseous nebula, and a cool dust shell \citep{zh91}.
The hot stellar component is assumed to be a blackbody of temperature $T_{*}$.
The gaseous nebular continuum is the sum of the free-free (f - f),
bound-free (b - f), and two-photon emissions using the emission coefficients given in \citet[p.~171]{kwok07} 
at an electron temperature $T_{e}$= 12\,000 K and an electron density $n_{e}$ = 7\,200 cm$^{-3}$ \citep{shaw89}. The emerging flux is therefore given by:

\begin{equation}
F_\lambda(total)=F_\lambda(s)+F_\lambda(g)+F_\lambda(d)
\end{equation}
where $F_\lambda(s), F_\lambda(g)$, and $F_\lambda(d)$ are the flux densities of the photospheric continuum
of the central star, the nebular continuum, and the dust thermal emission continuum,
respectively. 
The stellar flux is then given by

\begin{equation}
F_\lambda(s)=(\pi {\theta}_{*}^{2})B_\lambda(T_{*})
\end{equation}
where $\theta_{*}$, $T_{*}$  are respectively the angular radii and the effective temperatures of the central star, and $B_\lambda$($T_{*}$)  is the Planck functions for temperatures $T_{*}$. 
Given the uncertain extinction corrections for the visible photometry of the central star, and the fact that the B and V photometry lie in the Rayleigh-Jeans side of the blackbody,  the value of $T_*$ cannot be determined precisely.  Our best estimate for  
$T_*$ is $119\,000\pm 8500$ K, which is consistent with the previously reported values of 106\,000 K \citep{shaw89} and 115\,000 K \citep{zh91}.  With an assumed distance of 2.5 kpc \citep{cahn92}, the derived luminosity of the central star is $\sim$ 4600 L$_{\sun}$.  
According to the model of \citet{sch81}, the above temperature and luminosity of NGC 6644 imply an evolutionary age $>$6000 years.  The value is at least two times higher than the observed kinematic age (section \ref{lobes}).  This may suggest that the visible nebulosity (from which we estimate the kinematic age) originates from very late stellar winds, and observations with higher sensitivity are required to detect the fainter extended structures.

The near-IR photometry points suggest that there is excess emission between 2 and 5 $\mu$m which could be due to a cool companion, a hot dust component, or scattering from an unseen disk.  We have therefore added a blackbody of temperature 2950 K to the model fit.  Using the same assumed distance, this companion would have a luminosity of 14 L$_\sun$.

We find that the observed dust emission component is too broad to be fitted by a single blackbody. We therefore fit the  dust component by blackbodies of two different temperatures, a warm  ($T_{wd}$) and a cold  ($T_{cd}$) dust components. 
The total dust thermal emission is therefore given by the sum of two terms:

\begin{equation}
F_\lambda(d)=F_\lambda(wd)+F_\lambda(cd)=\frac{3M_{wd} Q_\lambda B_\lambda(T_{wd})}{4a \rho_{s} D^2} + 
\frac{3M_{cd} Q_\lambda B_\lambda(T_{cd})}{4a \rho_{s} D^2}
\end{equation}
where $M_{wd}$ and $M_{cd}$ are the masses of warm and cold dust components, $a$ is the grain radius which  depends on its physical shape, $\rho_{s}$ is the density of grain, $Q_\lambda$=Q$_{0}$
($\lambda$/$\lambda_{0}$)$^{-\alpha}$ is the grain emissivity function, and $D$ is the distance to the nebula \citep[p.~312]{kwok07}. According to our fit, the temperatures of the two dust components are 292 K and 123 K.
Assuming $Q$ (1 $\micron$)=0.1, $\rho_{s}$=1 g cm$^{-3}$, $a$=0.1 $\micron$, $\alpha$ =1, and $D$=2.5 kpc, we obtain 8 $\times$ 10$^{-8}$ M$_{\sun}$ and 3.12 $\times$ 10$^{-5}$ M$_{\sun}$ for the mass of the warm and cold dust components respectively.  
Without spectral coverage between 3 to 10 $\mu$m, it is difficult to constrain the temperature of the warm dust component precisely and the above mass estimate for the warm component is therefore subject to a large margin of error.

\begin{figure}
\plotone{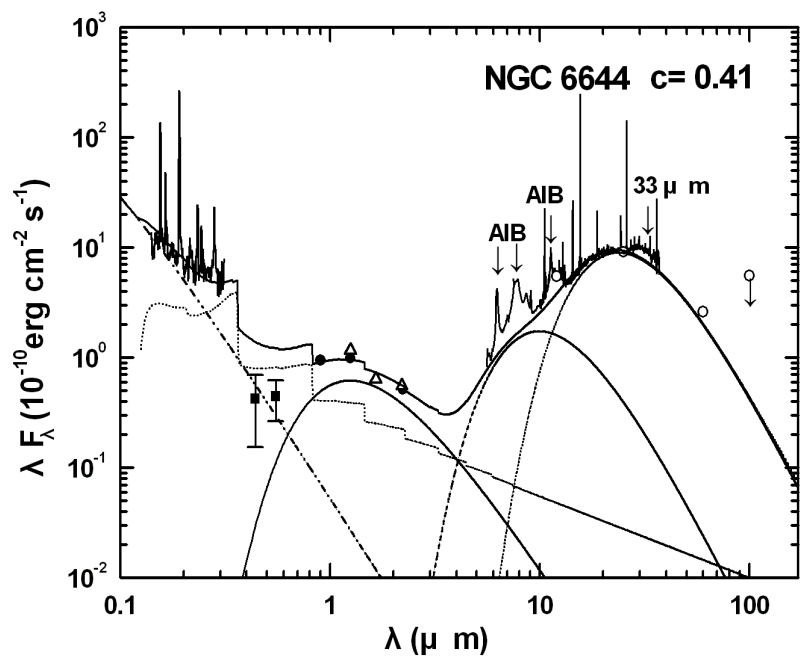}
\caption{SED of NGC 6644 in the wavelength range from 0.1 $\micron$ to 170 $\micron$. 
The filled squares are B and V photometry of the central star, the filled-circles are DENIS near-IR photometry, 
the open triangles are 2MASS photometry, and open circles are IRAS 12, 25, 60, and 100 $\micron$ photometry. 
Note that the flux measured from IRAS 100 $\micron$ is an upper limit. The four BB-like curves (from left to right) represent the central star, cool companion, and the two dust components.  The model curve with step jumps represents the nebular component.  The total flux from all components are plotted as a solid line on top. \label{sed}}
\end{figure}

The SED of the nebula corrected for extinction is shown in Fig.~\ref{sed}. We can see that the model gives a reasonable fit to the observed data from UV
through the far-infrared.  
From the total observed nebular flux of 5.3 $\times$ 10$^{-10}$ erg cm$^{-2}$ s$^{-1}$, an emission measure 
of 2.5 $\times$ 10$^{59}$ cm$^{-3}$ was derived. At a distance of 2.5 kpc and an electron density of 7200 
cm$^{-3}$, the total ionized gas mass is derived as 0.03 M$_{\sun}$.    From the model fitting, the fraction of total fluxes between 0.1 and 100 $\mu$m from the central stellar, nebular gaseous continuum emission, 
and dust components are 38$\%$, 18$\%$, and 44$\%$, respectively.  The stellar flux shortward of 0.1 $\mu$m is assumed to have been absorbed by the gas component.  

From the SED, we can see that there is a broad 30 $\micron$ feature in NGC 6644. 
This unidentified emission feature is commonly seen in carbon stars \citep{volk00}, post-AGB stars \citep{sz99, hri00, volk02} and young PNs.

\section{Simulated 3-D structure of NGC 6644}

The observations of NGC 6644 as reported in section \ref{lobes} suggest that NGC 6644 is a new member in the club of multipolar nebulae.  The class of multipolar nebulae, first described by  \citet{man96} and \citet{sahai98}, now include many prominent members including NGC 2440 \citep{lop98}, He 2-47 and M1-37 \citep{sahai00b}, He 2-113 \citep{sahai00c}, NGC 6881 \citep{su05}, NGC 6072 \citep{kwok10}, etc.  Multipolar nebulae, having a point-symmetric rather than axial symmetric structure, need 3-D modelling to visualize their structure.
With this aim, we have constructed a schematic model of NGC 6644 assuming three pairs of bipolar lobes and an equatorial dust torus using the software program SHAPE \citep{Steffen06}.

SHAPE is a morpho-kinematic modeling tool intended for the analysis of the 3D geometry and kinematic structure of gaseous nebulae. SHAPE uses both kinematic and spatial observations as a guide for the interactive reproduction of model nebulae. Once a 3D model has been constructed, parameters such as position angle and inclination can be manipulated in order to study the object from different orientations.  
In this work we are interested in studying the complex 3D morphology of multi-polar PN for which SHAPE is an ideal tool. We do not attempt to determine physical parameters such as the temperature and density structure and thus we do not perform any radiative transfer or hydrodynamic simulations. Instead, the brightness and location of each component is assigned based on the qualitative analysis of the observed images.

In addition to the observed double bipolar lobes $a$ and $b$, we hypothesize an additional pair of lobes that is perpendicular to the observed ionized torus in Fig.~\ref{hst}.  This pair of lobes will be referred to as lobe $c$.
We further assume that all three pairs of lobes are equal in length and  that the $b$ lobes are oriented in the plane of the sky (inclination 0$^\circ$).  
With these assumptions, we can then derive an inclination of 31$^\circ$ and -85$^\circ$ for the $a$ and $c$ lobes respectively. The estimated symmetry axes of these three lobes are along PA= -42\degr, -28\degr\ and 76\degr\ for  the $a, b$ and $c$ lobes respectively.

\begin{figure}
\epsscale{0.65}
\plotone{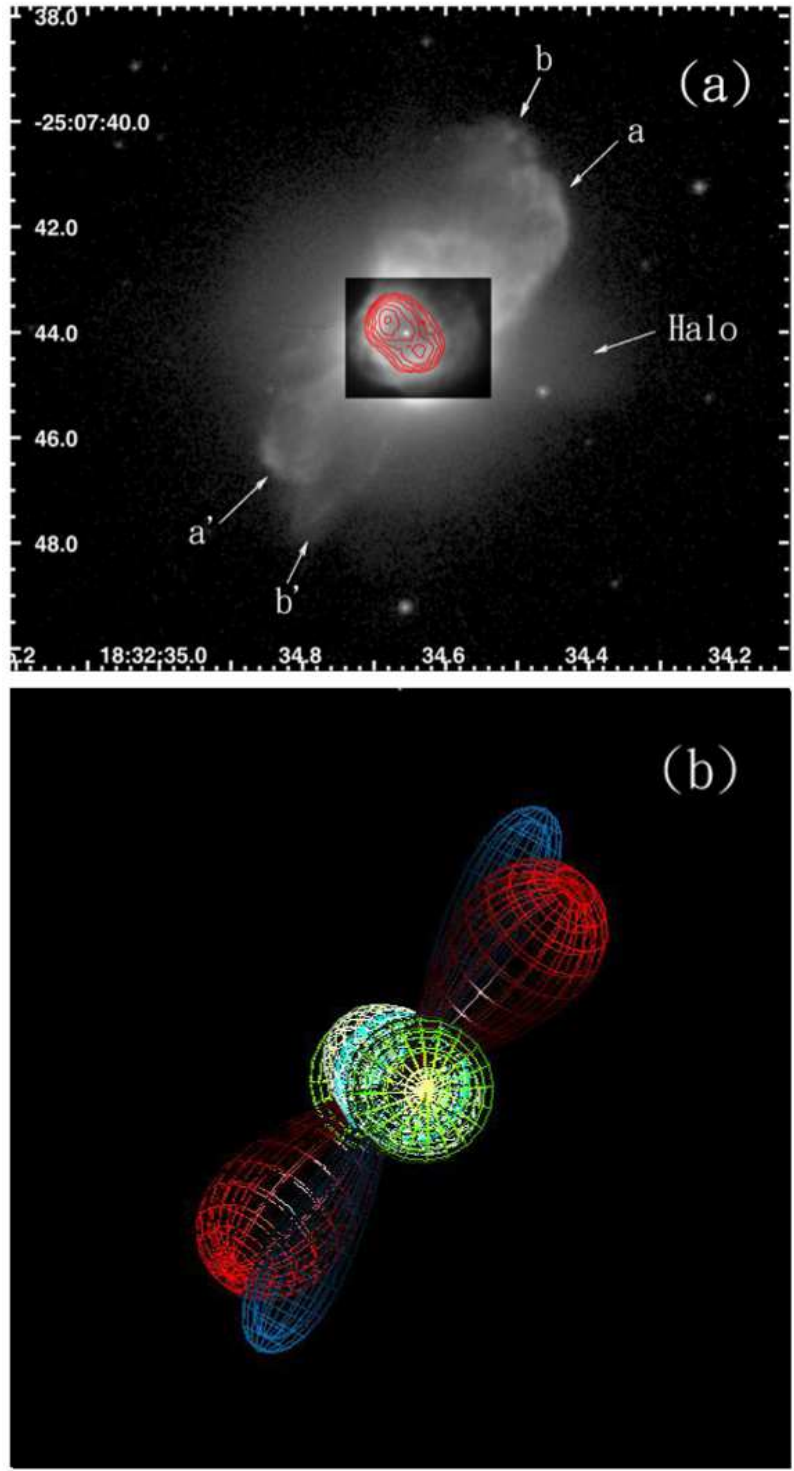}
\caption{Comparison of optical-infrared observations and the three-dimensional model of NGC 6644. 
(a) The gray frame shows the {\it HST} H$\alpha$ image, as in Figure 1. Linear contour plot of the T-ReCS PAH-2 
11.3 $\micron$ image was overlaid on the {\it HST} image (red lines). (b) SHAPE three-dimensional 
mesh model. The bipolar lobes $a$, lobe $b$, and lobe $c$ are displayed in red, blue, and green, 
respectively. The equatorial ring observed in the {\it HST} H$\alpha$ image is shown in light blue and the infrared torus observed in Gemini PAH2 image is in white. \label{model}}
\end{figure}

We created a dense torus as the counterpart of the infrared torus seen in the Gemini Si5 and PAH2 images with a center that coincides
with that of the three lobes. If we assume that the observed elliptical torus is a tilted circular disk,
then the model's major-to-minor axis ratio implies an inclination angle of 35$^\circ$ (with 90$^\circ$
being pole on). This value is in reasonable agreement with our previous estimate of 32$^\circ$ based on the ellipticity of the torus (see $\S$ 3.2).  In this model, the symmetry axis of lobe $a-a^\prime$ is perpendicular to the infrared dust torus.
A counterpart of the ring seen in the {\it HST} H$\alpha$ image and the Gemini [S~{\sc iv}] image is also included.
The model parameters of these structures are listed in Table 6 and a schematic mesh model is shown in Fig.~\ref{model}.

\begin{figure}
\epsscale{0.8}
\plotone{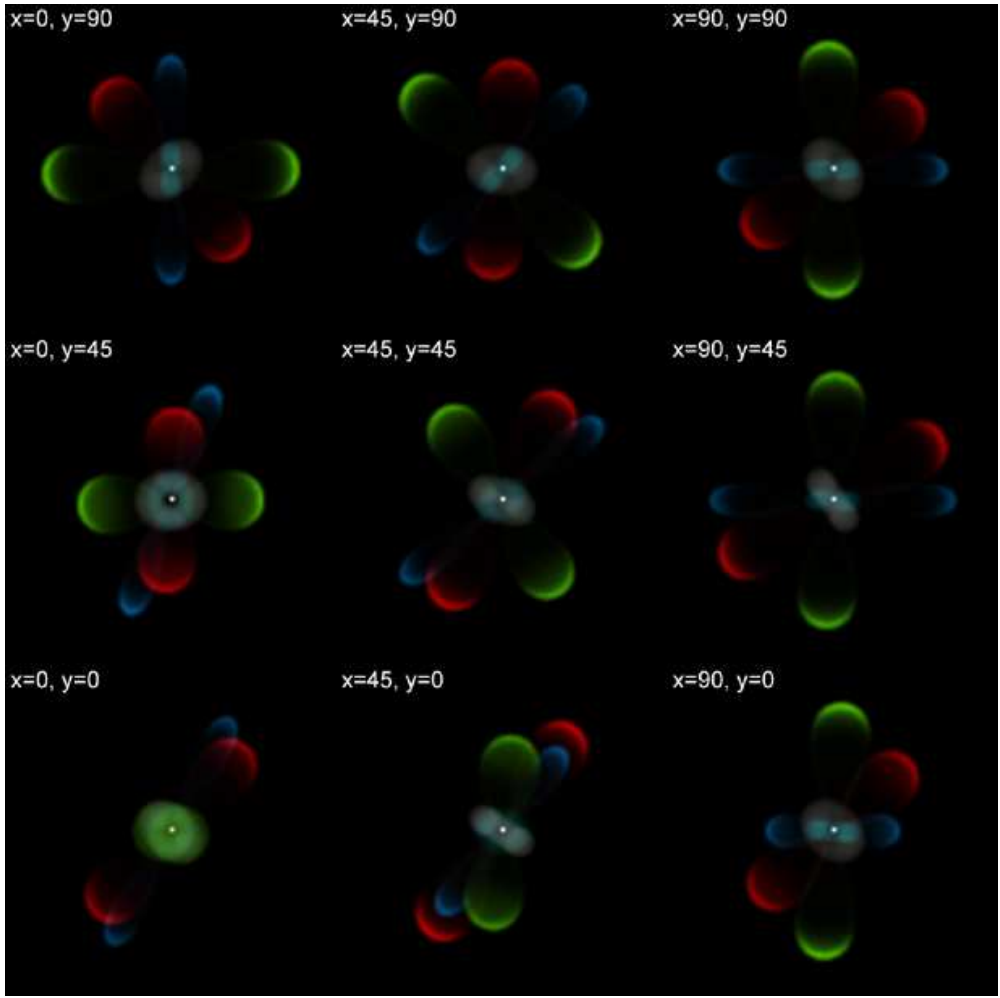}
\caption{The model of NGC 6644 as viewed from different angles. The rotation angles around the $x$ 
and $y$ axis (of the image plane) are given in the top corner of each image. The bottom left image represents the model as viewed from Earth (i.e. no rotation). The color scheme is the same as in Fig.~\ref{model}.  \label{shape}}
\end{figure}

This simple model can provide insight into what multipolar PNs look like at different orientations.
In Fig.~\ref{shape} we present a three-dimensional representation of NGC 6644 viewed from different angles.
The bottom left image is the model as seen from Earth, and the orientations of the other images are
as labeled. The $x$ and $y$ axes indicated are those of the image plane and the values represent rotations
around those axes in degrees.

When $y$ = 90$^\circ$, the nebula reveals three distinct pairs of lobes with an elliptical disk. This form
is similar to that found in two young PNs, He 2-47 and M 1-37 \citep{sahai00b}. When $y$ = 0$^\circ$ the bipolar
lobes appear closer together and overlap to varying extents. This type of structure is seen, for example,
in NGC 6072 \citep{kwok10} and the Frosty Leo Nebula \citep{sahai00}. 
The present model illustrates the possible complex 3-D structures that may be present in other objects and indeed it would be difficult to visualize the structure of NGC 6644 without the aid of such a model.
The main purpose of this simulation is to show how complex the structure of a multipolar PN can be and not as a quantitative fit to the observations.  A proper model of a complex object such as NGC 6644 will require kinematic information from all of the morphological features, which can be provided by integral field spectroscopy with large optical telescopes.

\section{Conclusions}

Although NGC 6644 appears to be a typical elliptical nebula, high angular resolution and dynamical range optical observations reveal that it in fact has an extremely complicated structure.
From {\it HST} and {\it Gemini} observations, we have identified at least two pairs of bipolar lobes and an infrared dust torus.  The orientation of the dust torus is approximately perpendicular to one pair of the lobes.  In addition, there is a ring of ionized gas lying almost on the plane of the sky.  We suggest that there could be a 3rd pair of bipolar lobes whose axis is perpendicular to this ring.
Several other faint sublobes at different radial distances and angular directions can also be seen.

It is clear that NGC 6644 is a multipolar nebula.  In order to illustrate its possible 3-D structure, we have constructed a model to show what it may look like when viewed from different perspectives.  Its simulated rotated images resemble some other multi-polar nebulae, suggesting that nebulae with different appearances may have similar intrinsic structures.

The emergence of the class of multipolar nebulae has greatly altered our perception of the morphological structures of PNs.  The commonly assumed simple structures of PNs are  probably the result of inadequate sensitivity and spatial resolution imaging.  
Deep imaging of PNs is needed to reveal the true intrinsic structures of PNs.


{\flushleft \bf Acknowledgements~}

Some of the data presented in this paper were obtained from the Multimission Archive at the Space Telescope Science Institute (MAST). STScI is operated by the Association of Universities for Research in Astronomy, Inc., under NASA contract NAS5-26555. Support for MAST for non-HST data is provided by the NASA Office of Space Science via grant NAG5-7584 and by
other grants and contracts.
Other parts of this work is based on observations made with the Gemini Observatory, which is operated by the Association of Universities for Research in Astronomy, Inc., under a cooperative agreement with the NSF on behalf of the Gemini partnership: the National Science Foundation (United States), the Particle Physics and Astronomy Research Council (United Kingdom), the National Research Council (Canada), CONICYT (Chile), the Australian Research Council (Australia), CNPq (Brazil) and CONICET (Argentina); and the {\it Spitzer Space Telescope}, which is operated by the Jet Propulsion Laboratory, California Institute of Technology under a contract with NASA. 
This work was partially supported by the Research Grants Council of the Hong Kong Special Administrative Region, China (project no. HKU 7031/10P.). 
NK acknowledges support by the Natural Sciences and Engineering Council of Canada, Alberta Ingenuity, and the Killam Trusts.

\clearpage

\clearpage

\begin{deluxetable}{clccc}
\tabletypesize{\scriptsize} \tablecaption{Summary of T-ReCS Observations of NGC 6644} \tablewidth{0pt}
\tablehead{\colhead{Observation Date} & \colhead{Filter Name} & \colhead{$\lambda_{c}$} &
\colhead{$\triangle$ $\lambda$} & \colhead{Exposures} \\
\colhead{} & \colhead{} & \colhead{($\micron$)} & \colhead{($\micron$)} & \colhead{(s)}}
\startdata
2007 Oct 10 & [SIV]-10.52 $\micron$ & 10.52 & 0.17 & 455.1  \\
2007 Oct 10 & PAH2-11.3 $\micron$ & 11.30 & 0.61 & 461.5  \\
2007 Oct 10 & Si5-11.66 $\micron$ & 11.66 & 1.13 & 238.9 \\
2007 Oct 10 & $Q_a$-18.30 $\micron$ & 18.30 & 1.38 &  238.9 \\
\enddata

\end{deluxetable}

\begin{deluxetable}{llcc}
\tabletypesize{\scriptsize} \tablecaption{Measured emission line fluxes in the IRS spectrum} \tablewidth{0pt}
\tablehead{\colhead{Wavelength} & \colhead{Identification} & \colhead{Observed flux} &
\colhead{Normalized flux} \\
\colhead{($\micron$)} & \colhead{} & \colhead{(10$^{-10}$ergs cm$^{-2}$s$^{-1}$)} & 
\colhead{(10$^{-10}$ergs cm$^{-2}$s$^{-1}$)}}
\startdata
10.51 & [S IV] & 0.1098 & 0.1109 \\
11.20 & AIB & 0.1513 & 0.1530 \\ 
11.76 & [Cl IV] & 0.0012 & 0.0013 \\
12.37 & HI (7-6) & 0.0024 & 0.0025  \\
12.81 & [Ne II] & 0.0055 & 0.0056  \\
13.09 & [Ar V] & 0.0019 & 0.0020 \\
14.32 & [Ne V] & 0.0225 & 0.0227 \\
15.55 & [Ne III] & 0.2366 & 0.2389  \\
18.71 & [S III] & 0.0157 & 0.0159 \\
19.07 & HI (8-7) & 0.0016 & 0.0017 \\
24.32 & [Ne V] & 0.0087 & 0.0088  \\
25.88 & [O IV] & 0.1440 & 0.1454 \\
33.47 & [S III] & 0.0038 & 0.0039 \\
36.01 & [Ne III] & 0.0198 & 0.0200 \\
\enddata

\end{deluxetable}

\clearpage

\begin{deluxetable}{lcccc}
\tabletypesize{\scriptsize} \tablecaption{Comparison of IRS and Gemini T-ReCS Observations for NGC 6644} 
\tablewidth{0pt}
\tablehead{\colhead{Filter Name} & \colhead{Central Wavelength $\lambda_{c}$} 
& \colhead{Band Width $\triangle$ $\lambda$} & \colhead{Flux Density} & \colhead{IRS Flux Density$^a$} \\
\colhead{} & \colhead{($\micron$)} & \colhead{($\micron$)} & \colhead{(Jy)} & \colhead{(Jy)}}
\startdata
SIV-10.52 $\micron$ & 10.52 & 0.17  & 4.18 & 4.29 \\
PAH2-11.3 $\micron$ & 11.30 & 0.61 & 2.31 & 2.41 \\
Si5-11.66 $\micron$ & 11.66 & 1.13 & 2.39 & 2.32 \\
$Q_a$-18.30 $\micron$ & 18.30 & 1.38 & 5.78 & 5.01 \\
\enddata

\tablenotetext{{\it a}}{The IRS flux density is obtained from integrating the convolution of filter
transmission curve and Spitzer IRS spectrum.}

\end{deluxetable}

\begin{deluxetable}{clcc}
\tabletypesize{\scriptsize} \tablecaption{Available IUE data and IRS observation} \tablewidth{0pt}
\tablehead{\colhead{PN} & \colhead{Name} & \colhead{Instrument} & \colhead{Exposures} \\
\hline
& & \multicolumn{2}{c}{IUE Spectra}}
\startdata
G 008.3-07.3 & NGC 6644 & SWP 31711 & 1800  \\
 & & SWP 01734 & 1800 \\
 & & LWR 01630 & 2400 \\ 
\hline
& & \multicolumn{2}{c}{IRS Observation} \\
\hline
G 008.3-07.3 & NGC 6644 & AORkey 11334400 & 347  \\
\enddata

\end{deluxetable}

\clearpage

\begin{deluxetable}{ll}
\tabletypesize{\scriptsize}
\tablecaption{Other photometric measurements of NGC 6644}
\tablewidth{0pt}
\tablehead{
\colhead{Filters} & \colhead{Flux/Flux density} \\
\hline
\multicolumn{2}{c}{Central star}
}
\startdata
B (mag)$^a$ & 16.6  \\
V (mag)$^a$ & 15.63  \\
-log F (H$\beta$) (ergs.cm$^{-2}$.s$^{-1}$)$^a$ & 11.01 \\
\hline
 \multicolumn{2}{c}{Nebula} \\
\hline
DENIS I$^b$ (mag) & 12.678 \\
DENIS J$^b$ (mag) & 11.743  \\
DENIS K$^b$ (mag) & 10.695  \\
2MASS J$^c$ (mag) & 11.562 \\
2MASS H$^c$ (mag) & 11.599  \\
2MASS Ks$^c$ (mag) & 10.627  \\
IRAS F 12 $\micron$$^d$ (Jy) & 2.19 \\
IRAS F 25 $\micron$$^d$ (Jy) & 7.57 \\
IRAS F 60 $\micron$$^d$ (Jy) & 4.22 \\
IRAS F 100 $\micron$$^d$ (Jy) & 18.39  \\
\enddata
\tablenotetext{{\it a}}{From Shaw \& Kaler (1989);}
\tablenotetext{{\it b}}{From DENIS database;}
\tablenotetext{{\it c}}{From Ramos-Larios \& Phillips (2005); }
\tablenotetext{{\it d}}{From Tajitsu \& Tamura (1998), the color- and diameter-corrected
IRAS fluxes were given. Note that the flux measured from 100 $\micron$ is an upper limit.}

\end{deluxetable}

\begin{deluxetable}{ccccc}
\tabletypesize{\scriptsize} \tablecaption{Comparison of Observed and Model Parameters of the Lobes and Torus} \tablewidth{0pt}
\tablehead{
 & & \multicolumn{2}{c}{Value} & \\
\cline{3-4} 
\colhead{Parameters} & \colhead{Observed} & \colhead{Model} & \colhead{Comment}} 
\startdata
Position angle of lobe $a-a^\prime$ &  -43$^\circ\pm$2$^\circ$ & -42$^\circ$ & \nodata \\
Position angle of lobe $b-b^\prime$ &  -27$^\circ\pm$2$^\circ$ & -28$^\circ$ & \nodata \\
Position angle of lobe $c-c^\prime$ &  \nodata & 76$^\circ$ & \nodata \\
Inclination angle of lobe a-a' & \nodata & 31$^\circ$ & Assuming the orientation angle of sky plane 
is 0$^\circ$ \\
Inclination angle of lobe $b-b^\prime$ & \nodata & 0$^\circ$ & " \\
Inclination angle of lobe $c-c^\prime$ & \nodata & -85$^\circ$ & " \\
Position angle of dust torus & -44$^\circ\pm$3$^\circ$ & -41$^\circ$ & Derived from peak position of torus \\
Inclination angle of dust torus  & 32$^\circ$ & 35$^\circ$ & Derived from the major-minor axis ratio \\ 
Position angle of ionized torus & $-55^\circ\pm$3$^\circ$ & -60$^\circ$ & \\
Inclination angle of ionized torus  & -106$^\circ$ & -100$^\circ$ &  \\ 
\enddata

\end{deluxetable}


\begin{thebibliography}{}

\bibitem[Aller \& Keyes(1988)]{aller88}
Aller, L. H. \& Keyes, C.D. 1988, PASP, 100, 192



\bibitem[Beaulieu et al.(1999)]{bea99} 
Beaulieu, S. F., Dopita, M. A., \& Freeman, K. C., 1999, \apj, 515, 610


\bibitem[Cahn et al.(1992)]{cahn92} 
Cahn, J. H., Kaler, J. B., \& Stanghellini, L., 1992, \aaps, 94, 399

\bibitem[Cohen et al.(1999)]{cohen99} 
Cohen, M., Walker, R. G., Carter, B., et al., 1999, \aj, 117, 1864

Garc\'{i}a-Segura, G., 1997, \apj, 489, L189


\bibitem[G\'{o}ny et al.(2004)]{gorny04} 
G\'{o}rny, S. K., Stasi\'{n}ska, G., Escudero, A. V., et al., 2004, \aap, 427, 231



\bibitem[Higdon et al.(2004)]{higdon04} 
Higdon, S. J. U., Devost, D., Higdon, J. L., et al., 2004, \pasp, 116, 975
\bibitem[Houck et al.(2004)]{houck04} 
Houck, J. R., Appelton, P. N., Armus, L., et al., 2004, \apjs, 154, 18
\bibitem[Hrivnak, Volk, \& Kwok(2000)]{hri00} 
Hrivnak, B. J., Volk, K., \& Kwok, S., 2000, \apj, 535, 275

\bibitem[Hubble(1921)]{hubble21} 
Hubble, E., 1921, PASP, 33, 175




\bibitem[Kwok(2007)]{kwok07} 
Kwok, S., 2007, {\it Physics and Chemistry of the Interstellar Medium}, (Sausalito, CA:Univ. 
Science Books)




\bibitem[Kwok \& Su(2005)]{su05}
Kwok, S., \& Su, K.Y.L. 2005, ApJ, 635, L52

\bibitem[Kwok, Volk, \& Hrivnak(1999)]{kwok99} 
Kwok, S., Volk, K., \& Hrivnak, B. J., 1999, \aaps, 350, 35 

\bibitem[Kwok, Volk, \& Bernath(2001)]{kwok01}
Kwok, S., Volk, K., \& Bernath, P. 2001, \apj, 554, L87

\bibitem[Kwok et al.(2010)]{kwok10} 
Kwok, S., Chong, S. N., Hsia, C. H., et al., 2010, \apj, 780, 93 


\bibitem[L\'opez et al.(1998)]{lop98}
L\'opez, J. A., Meaburn, J., Bryce, M., \& Holloway, A. J. 1998, ApJ, 493, 803



\bibitem[Manchado, Stanghellini, \& Guerrero(1996)Manchado et al.]{man96}
Manchado, A., Stanghellini, L., \& Guerrero, M.A. 1996, ApJ, 466, L95



\bibitem[Perea-Calder\'{o}n et al.(2009)]{perea09} 
Perea-Calder\'{o}n, J. V., Garcia-Hern$\acute{a}$ndez, D. A., Garcia-Lario, P., Szczerba, R., Bobrowsky, M. 2009, \aap, 495, 5
\bibitem[Ramos-Larios \& Phillips(2005)]{ramos2005} 
Ramos-Larios G., \& Phillips J. P., 2005, \mnras, 357, 732

\bibitem[Sahai(2000a)]{sahai00} 
Sahai, R., 2000a, Asymmetrical Planetary Nebulae II: From Origins to Microstructures, ASP Conference Series, Vol. 199. Eds. J. H. Kastner, N. Soker, and S. Rappaport,  p. 209

\bibitem[Sahai(2000b)]{sahai00b} 
Sahai, R., 2000b, \apj, 537, L43

\bibitem[Sahai \& Trauger(1998)]{sahai98}
Sahai, R., \& Trauger, J.T. 1998, AJ, 116, 1357

\bibitem[Sahai et al.(2000c)]{sahai00c}
Sahai, R., Nyman, L., Wootten, A. 2000c \apj, 543, 880

\bibitem[Sch\"onberner(1981)]{sch81}
Sch\"onberner, D. 1981, A\&A, 103, 119



\bibitem[Shaw \& Kaler(1989)]{shaw89} 
Shaw, R. A., \& Kaler, J. B., 1989, \apjs, 69, 495



%

\bibitem[Stanghellini et al.(1993)]{stan93} 
Stanghellini, L., Corradi, R. L. M., \& Schwarz, H. E., 1993, \aap, 279, 521

\bibitem[Steffen \& L\'opez(2006)]{Steffen06} 
Steffen, W., \& L\'opez, J. A., 2006, \rmxaa, 42, 99


\bibitem[Szczerba et al.(1999)]{sz99} 
Szczerba, R., Henning, Th., Volk, K., Kwok, S., Cox, P.  1999, \aap, 345, L39  

\bibitem[Tajitsu \& Tamura(1998)]{tajitsu98} 
Tajitsu, A., \& Tamura, S., 1998, \aj, 115, 1989


\bibitem[Volk, Xiong, \& Kwok(2000)]{volk00} 
Volk, K., Xiong, G. Z., \& Kwok, S., 2000, \apj, 530, 408

\bibitem[Volk et al.(2002)]{volk02} 
Volk, K., Kwok, S., Hrivnak, B., Szczerba, R.  2002, \apj, 567, 412

\bibitem[Zhang \& Kwok(1991)]{zh91} 
Zhang, C. Y, \& Kwok, S., 1991, \aap, 250, 179

\end{thebibliography}
\end{document}